\documentstyle[prl,aps,epsfig,amssymb]{revtex}
\begin{document}
\draft
\title{Electrodynamics of metallic photonic crystals  and  problem of
left-handed materials}
\author{A. L. Pokrovsky and A. L. Efros}
\address{\vspace{-0.1cm} Department of Physics, 
	University of Utah, Salt Lake City UT, 84112 USA}
\maketitle

\begin{abstract}

An analytical theory of low frequency electromagnetic waves 
in metallic photonic crystals with a small 
volume fraction of a metal is presented.
The evidence for such waves has been obtained recently 
by experiments and computations.
The cutoff frequency of these waves, $\omega_0$ is studied.
An analytical expression for 
the permittivity  $\epsilon$ is obtained and shown 
to be negative below $\omega_0$.  
If the crystal is embedded
into a medium with a negative $\mu$, there are no
propagating modes at any frequency. 
Thus, such a compound  system is not a left handed material (LHM). 
The recent experimental  results on the LHM are discussed.

\end{abstract}


In his seminal work Veselago\cite{ve} has shown that if in some
frequency range
both the permittivity $\epsilon$ and permeability $\mu$ are negative,
the electromagnetic waves (EMW's) propagate
but they have some peculiar properties. All these properties 
come from the fact that vectors {\bf k, E, H} form a left
handed rather than a right handed set.
It follows that the Poynting vector
and the wave vector {\bf k} have opposite directions.
The materials with these properties are called the 
left handed materials (LHM's).

The idea that a  metallic photonic crystal (MPC) may be a technological 
base for the LHM \cite{p2,sm}
appears as a result of the computational and experimental studies of a few
 groups\cite{yab,soc1,soc,p3,sm2} which have found the EMW's 
in the MPC propagating above a
very low cutoff frequency. 
The MPC's they have considered are three- or two-dimensional 
lattices of thin straight metallic
wires.
Their discovery is very interesting and important because the waves
propagate under the condition $f\sigma/\epsilon_0 \omega \gg 1$,
where $f \sigma$ is an average conductivity, $f$ being the volume fraction
 of a metal in the system. 
This propagation must be due to the MPC structure,
 because it would be impossible in a 
homogeneous medium with the conductivity $f\sigma$.

Various groups obtained different cutoff frequencies and they understood 
them differently.
The group of Soukoulis\cite{soc} qualitatively interpreted the effect 
of propagation in
 terms of waveguide modes, while the group of Pendry\cite{p3}
 presented a completely original physical picture based upon a new
  longitudinal mode, called ``plasma mode".
According to  Pendry {\it et al.}\cite{p3} 
the resulting permittivity has a plasma-like behavior
\begin{equation}
\label{eps}
\frac{\epsilon}{\epsilon_0}= 1-{\omega_p^2\over
\omega(\omega+i\Gamma)},
\end{equation}
however, the ``plasma" frequency contains the light velocity and has a form
\begin{equation}
\label{om}
\omega_p^2={2 \pi c^2 \over d^2 \ln (d/R)},
\Gamma=\frac{\epsilon_0d^2\omega_p^2}{\pi R^2\sigma},
\end{equation}
where $d$ is the lattice constant, $R$ is the 
radius of the metallic wires,
$\sigma$ is the static conductivity of the metal. The same results for
$\epsilon$ and $\omega_p$ have been later obtained 
theoretically by Sarychev and Shalaev\cite{sar}.

The San Diego group has accepted the ``plasma model" and considered the
negative $\epsilon$ at $\omega<\omega_p$ as one of the two crucial 
conditions for creation of the LHM. 
To obtain negative $\mu$  
the split ring
resonators (SRR's) are added to the MPC\cite{sm,sm3,p1}. 
The first observation of the negative refraction
at the interface of this compound system and vacuum 
has been reported recently\cite{sm3}.
The negative refraction is the most important manifestation of the LHM.

We claim in this paper that the ``plasma mode" in the MPC is in fact 
an EMW and that is why the experiments
of the San Diego group cannot be simply understood 
using $\epsilon$ of the MPC and $\mu$ of the SRR's. 
Our paper is organized as follows. 
First we discuss the arguments of Pendry {\it et al.}\cite{p3} in favor
of the plasma model and the 
theoretical approach by Smith {\it et al.} \cite{sm2}. 
Then we derive and solve 
an exact dispersion equation for the cutoff frequency $\omega_0$ and
find $\epsilon(\omega)$. The results
are different from Eqs.(\ref{eps},\ref{om}) but $\omega_0$ is  in a very
good agreement with all
computational and experimental data we are aware of.
It follows that all groups discuss the same mode. 
The permittivity becomes negative at $\omega<\omega_0$.
Then  we show that
the MPC does not support any waves if it is embedded in the medium
with negative $\mu$ and
discuss this result together with the experimental results of the San
Diego group.

The Pendry group has proposed that the plasma mode in the MPC
appears as a
result of an electromagnetic renormalization of the electron mass in the
second order in $1/c^2$.  Due to this renormalization the electron
mass becomes $\approx 15 $ times larger than the mass of a proton. As
a result, the plasma frequency of the metal shifts down into the GHz
range.  It is known, however,
that the fields can be excluded from the interaction energy and the
Lagrangian can be written as a function of instantaneous velocities
and coordinates of the interacting charges keeping terms of the order of
$c^{-2}$\cite{jac}.  This so-called Darwin Lagrangian does not contain
any mass renormalization due to the interaction. It is also strange
that one should care about both mass renormalization and plasma
frequency in connection with a problem based upon
Maxwell equations with a static conductivity.
These equations contain neither electron mass nor plasma frequency,
and the very concept of electrons is not important for them.

The interpretation of the electric permittivity
[Eqs.(\ref{eps},\ref{om})]
as given by Smith {\it et al.}\cite{sm2} is in terms of
conventional electrodynamics
and it provides a reasonable  basis for a discussion. 
As far as we understand, it is based upon the ansatz
\begin{equation}
\label{an}
E_{ac}= \overline{E} + i \pi R^2 \omega L j,
\end{equation}
where $E_{ac}$ is the electric field acting on a wire, $\overline{E}$
 is the average field, $j=\sigma E_{ac}$ is the current density, and $L$ is the
 self-inductance of the wire per unit length. Using the ansatz
 (\ref{an}) one can easily get Eqs.(\ref{eps},\ref{om}).

Now we derive an exact dispersion relation for the s-polarized
EMW  under the condition $f\ll 1$ 
in the system of infinite parallel thin straight wires
ordered in a square lattice.
The electric field of the wave is along
the wires (z-axis), while the wave vector ${\bf k}$ is in the $x$-$y$
plane.  Assume that the total current in each wire is $I_0\exp[
-i(\omega t -{\bf k\cdot r}_i)]$, where ${\bf r}_i$ is the
 two-dimensional radius-vector of the wire in the $x$-$y$ plane.  The
 external solution for electric field $E_z$ of one wire with ${\bf
 r}_i=0$ has a form
\begin{equation}
E_{z}^i= (I_0 \mu_0 \omega/4) H^{(2)}_0(\omega\rho/c),
\end{equation}
where $H^{(2)}_0=J_0-iN_0$ is the Hankel function which decays
exponentially at ${\rm Im}\omega <0$. Neither $J_0$ nor $N_0$ has
this important property.  Here and below we omit the time dependent factor.
The solution is written in cylindrical coordinates $z,\rho,\phi$ and
it obeys the boundary condition $ B_{\phi}=(i/ \omega)dE_z/ d\rho=
I_0\mu_0/ 2\pi \rho$ at $\rho = R$.

The electric field created by all wires is
\begin{equation}
\label{e}
E_z({\bf r})={I_0 \mu_0 \omega\over 4}e^{i{\bf k\cdot r}}
        \sum_{j}e^{i{{\bf k\cdot} ({\bf r}_j-{\bf r})}}
        H^{(2)}_0({\omega\over c} \rho_j),
\end{equation}
where $\rho_j = \sqrt{(x-x_j)^2+(y-y_j)^2}$,
summation is over all sites of the square lattice and the sum is
a periodic function of ${\bf r}$.

The dispersion equation follows from the boundary condition \cite{lan} 
that relates the total electric field at the surface of any wire $l$
to the total current through this wire
$E_{zl}=I_0 \exp{(i {\bf k}\cdot{\bf r}_l)}/\sigma_{ef} \pi R^2$, where
$\sigma_{ef}=2 \sigma J_1(\kappa R)/\kappa R J_0(\kappa R)$,
$\kappa=(1+i)/\delta$, and $\delta$
is the skin depth. At small frequencies, when $\delta > R$ one gets
 $\sigma_{ef}\approx \sigma$. At high frequencies, when $\delta \le
 0.1 R$, one gets the Rayleigh formula $\sigma_{ef}\approx (1+i)\sigma
 \delta/R$.

Note, that the EMW exists mostly if the skin-effect in the wires is
strong.  Using Eqs.(\ref{eps},\ref{om}) one can show that $\Gamma
/\omega_p = \delta^2 / (R^2 \ln{d/R})$, where $\delta$ is the skin
depth at $\omega = \omega_p$ (see also Ref.\cite{shal}).  We show
below that the exact solution has similar properties.  That is why we
mostly concentrate here on the case of the strong skin-effect.

Finally, the dispersion equation for
$\omega({\bf k}) \equiv (\chi - i \gamma)c/d$ has a form
\begin{equation}
\label{dsp}
(\chi - i \gamma) \sum_{l,m} e^{i d( k_x l+ k_y m)} H_0^{(2)} (z_{lm}) =
 \frac{4 c \epsilon_0}{d \sigma_{ef} f},
\end{equation}
where $z_{lm} =  (\chi - i \gamma) \sqrt{l^2 + m^2 + (R/d)^2}$, $l$ and $m$ 
are integer numbers,
the small term $(R/d)^2$ under the square root is important
 only when $l = m = 0$.  Taking real and imaginary parts of Eq.(\ref{dsp})
one gets two equations for $\chi$ and $\gamma$.

In the continuum approximation one can substitute the summation by 
integration in Eq.(\ref{dsp}) to get
\begin{equation}
\label{eqq}
1 - \frac{c^2 k^2}{\omega^2} +
 i \frac{f \sigma_{ef}}{\epsilon_0 \omega} = 0.
\end{equation}
Equation (\ref{eqq}) describes propagation of a plane wave
through a homogeneous medium with the conductivity $f \sigma_{ef}$,
which is possible if $f \sigma_{ef} /\epsilon_0 \omega \ll 1$.
However, outside the continuum approximation there are propagating
modes in the low frequency range $f \sigma_{ef} /\epsilon_0 \omega \gg 1$.
For these modes the fields are
strongly modulated inside the lattice cell and ${\bf E}_z ({\bf r})$ is
close to zero near
each wire so that the absorption is small.

\begin{figure}
\centering\epsfig{file=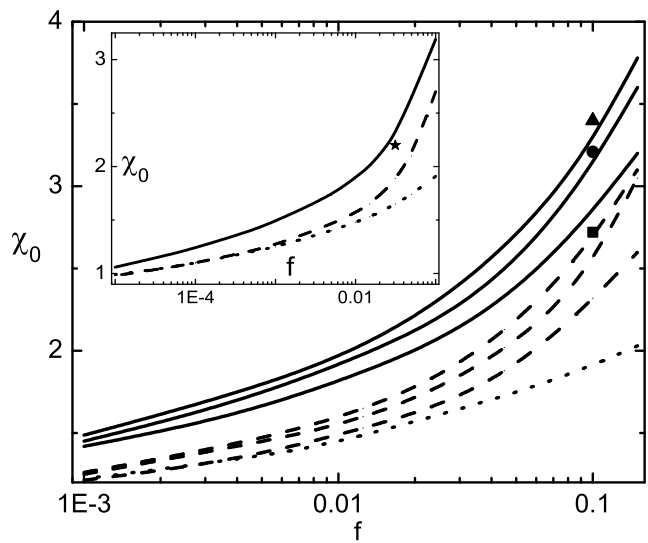, width=8.6cm}
\vspace{0.2cm}
\caption{The real part of the dimensionless cutoff frequency
        $\chi_0$ as a function of the
        volume fraction of metal $f$.
 	Solid, dashed, and dotted lines represent solutions
 	of Eq.(\ref{dsp0}), Eq.(\ref{dsp_a}), Eq.(\ref{om}) respectively.
 	The experimental data of Ref.\protect\cite{sm}($\bigstar$)
 	are shown together with numerical data of
 	Ref.\protect\cite{soc} ($\blacktriangle$,
 	$\bullet$, $\blacksquare$).
 	On the main plot
  	$\alpha = 0.024$, $d = 12.7 \mu m$ for upper solid, upper dashed
	lines,
  	and for the point  $\blacktriangle$;
  	$\alpha = 0.078$, $d = 1.27 \mu m$ for middle solid, middle
        dashed lines, and for the point $\bullet$;
        $\alpha = 0.246$, $d = 0.13 \mu m$ for lower solid,
 	lower dashed lines, and for the point $\blacksquare$.
        On the insert $\alpha = 3.4\cdot 10^{-4}$, $d = 8.0 mm$.}
\label{fig}
\end{figure}

We begin with the frequency $\omega_0$ which is the  solution of
Eq.(\ref{dsp})
at  $|{\bf k}| = 0$
\begin{equation}
\label{dsp0}
(\chi - i \gamma) \sum_{l,m} H_0^{(2)} (z_{lm}) =
 \frac{4 c \epsilon_0}{d \sigma_{ef} f}.
\end{equation}
This  mode
is an eigenmode of the system and  its frequency is the 
cutoff frequency for the EMW's.
The numerical results for the real part of the frequency
are shown in Fig. \ref{fig}.
One can see that $\chi$ is of the order of few units.
The values of $\gamma$ are of the order of the right hand side of 
Eq.(\ref{dsp0}).
Thus, $\chi \gg \gamma$ if $f \sigma_{ef} / \epsilon_0 \omega \gg 1$.

In addition to the numerical solution we propose an approximation valid
at very small $f$, when $|\ln f|\gg 1$.
We separate the term with $l = m = 0$ and
 substitute the rest of the sum by the integral.  Then
\begin{equation}
\label{dsp_a}
(\chi - i \gamma) \left[ \frac{2 i}{\pi} \left( \ln{\frac{2}{\chi
        \sqrt{f/\pi}}} - {\bf C} \right) - \frac{4 i}{(\chi - i
        \gamma)^2} \right] = \frac{4 c \epsilon_0}{d \sigma_{ef}(\chi)
        f},
\end{equation}
where we assume that $\gamma \ll \chi$.  
Here ${\bf C}$ is the Euler's constant. 
The second term at the left hand side represents the average
field $\overline{E}_z$ which can be found from the Maxwell equation
by the following way.
One can show that the average (or macroscopic) magnetic induction ${\bf
\overline{B}}$ is zero. Then $\nabla \times {\bf \overline{B}}=0$,
$\overline{\jmath}_z  + \epsilon_0 \partial
        \overline{E}_z / \partial t = 0$, and
\begin{equation}
\label{e_ave}
\overline{E}_z = \frac{I_0}{i \omega \epsilon_0 d^2}.
\end{equation}
The second term in the square brackets of Eq.(\ref{dsp_a}) is $4
\overline{E}_z d/\mu_0 I_0 c = -4 i/(\chi - i \gamma)$.  Thus, the
expression in the square brackets describes deviation of the field
acting on a wire  from the average field and
it is assumed that only one term of the sum in 
Eq.(\ref{dsp0}) makes this difference.
This approximation is similar to the ansatz
of Smith {\it et al.}\cite{sm2}.
The main difference between Eq.(\ref{an})
and Eq.(\ref{dsp_a}) which is crucial
for the imaginary part of the frequency and also important for the
real part, is the frequency dependence of $\sigma_{ef}$ due to the
skin effect.  Say, in the work by Smith {\it et al.}\cite{sm} $\delta /
R \approx 7 \cdot 10^{-4}$, so that the skin effect is very strong.

Figure \ref{fig} compares our result for the real part of the
cutoff frequency given by  Eq.(\ref{dsp0}) with the results given by
Eqs.(\ref{om},\ref{dsp_a}).  In these calculations we assume
that $\sigma_{ef}$ is given by the Rayleigh formula so that the right hand
side of Eqs.(\ref{dsp0},\ref{dsp_a}) has a form $\alpha (1-i)
\sqrt{\chi}$, where $\alpha = 2 c \epsilon_0 R/f d \sigma \delta$ and
$\delta$ is taken at $\omega = c/d$. One can see from Fig.\ref{fig} that
approximation Eq.(\ref{dsp_a}) is much better than approximation
Eq.(\ref{om}).  Both approximations coincide at small $f$ and are
accurate at extremely low values of $f$ ($\sim 10^{-7}$) when the
logarithmic term in Eqs.(\ref{om},\ref{dsp_a}) is very large.  The
computational and experimental data of Ref.\cite{soc,sm} are also
shown at Fig.~\ref{fig} and they are in a good agreement with
Eq.(\ref{dsp0}).
The results of Pendry group \cite{p3} (not shown) are also in a good
agreement with Eq.(\ref{dsp0}).
Thus, we can make the conclusion
that the San Diego group, group of Pendry and Soukoulis group
discuss the same mode but at different values of parameters  
and that our analytical theory describes the same mode as well.

Now we find the component of electric permittivity
$\epsilon(\omega) = \epsilon_{zz}$,
which describes the s-polarized extraordinary waves
in the uniaxial crystal.
It is defined by the relation
\begin{equation}
\epsilon = \epsilon_0 + i\frac{\tilde{\sigma} d}{\chi c},
\end{equation}
where effective macroscopic conductivity $\tilde{\sigma}$ relates
average current density $\overline{\jmath}_z$ to the average electric
field $\overline{E}_z$ by equation $\overline{\jmath}_z =
\tilde{\sigma} \overline{E}_z$.  To find $\tilde{\sigma}$ we introduce
an external electric field ${\cal E}_z e^{ -i \omega t}$.  The
average field $\overline{E}_z$ is given by equation
\begin{equation}
\label{ext_ave}
\overline{E}_z = {\cal E}_z - i \frac{I_0 \mu_0 c}{d (\chi - i
\gamma)},
\end{equation}
where the second term is the discussed above average field created by
the wires.  The boundary condition on a wire now has a form
\begin{equation}
\label{bc}
\frac{I_0 c (\chi - i \gamma) \mu_0}{4 d} \sum_{l,m}{}
      H_0^{(2)}\left(z_{lm}\right) + {\cal E}_z = \frac{I_0}{\pi R^2
      \sigma_{ef}}.
\end{equation}

Making use of Eqs.(\ref{ext_ave},\ref{bc}) one can find a
relation between the current and the average field which gives both
 $\tilde{\sigma}$ and $\epsilon$. Finally one gets
\begin{equation}
\label{epsilon}
\frac{\epsilon}{\epsilon_0} = 1-\left\{ \displaystyle \frac{\chi}
{\chi - i \gamma} - i \chi \left[ \frac{\chi - i \gamma}{4} \sum_{l,m}
H_0^{(2)}\left( z_{lm}\right) -\frac{c \epsilon_0}{f d \sigma_{eff}}
\right]\right\}^{-1}.
\end{equation}
Assuming that the medium is transparent ($\gamma \rightarrow 0$,
$\sigma \rightarrow \infty$) one can get the electric permittivity from
the total energy density $U$ of the electric field 
$U = (1/2) \overline{E_z^2} d(\omega \epsilon)/d \omega$.  We have
checked that this method gives the same result for $\epsilon(\omega)$.
The expression in the square brackets of Eq.(\ref{epsilon})
is the dispersion equation (\ref{dsp0}).
One can see that ${\rm Re}\epsilon$
changes sign at $\omega = \omega_0$ and becomes negative 
at $\omega<\omega_0$, where 
$\omega_0$ is the root of the dispersion equation (\ref{dsp0}).
The derivative $d ({\rm Re} \epsilon)/d \omega$ at $\omega = \omega_0$
is the same as the derivative which follows from Eq.(\ref{eps}) 
at $\omega = \omega_p$. However, far from the root of $\epsilon$ equations
(\ref{epsilon}) and (\ref{eps}) differ substantially.

To find $\omega({\bf k})$ one should solve  Eq.(\ref{dsp}).
For small $|{\bf k}|$ one can get analytical result 
$\omega^2 = c^2 k^2 + \omega_0^2$, which is isotropic in the 
$x$-$y$ plane.

Now we discuss the possibility 
of creation of the LHM using the negative $\epsilon$ of the MPC. 
Suppose that the wires are embedded into a medium with the
negative magnetic permeability $\mu$.  One can see that in this case
the propagation of any EMW is suppressed.  Indeed, instead of
Eq.(\ref{dsp}) one gets equation
\begin{equation}
\label{dsp_bad}
\sum_{j} e^{i ( k_x x_j+ k_y y_j)} K_0 \left( \frac{\omega}{c_0}
       \sqrt{x_j^2 + y_j^2 + R^2} \right) = \frac{2 i}{|\mu| \omega
       R^2 \sigma_{ef} },
\end{equation}
where $c_0=1/\sqrt{|\mu| \epsilon_0}$ and $K_0$ is the modified Bessel
function.  One can see that at $|{\bf k}| = 0$ all the terms on the
left hand side of this equation are positive and real if ${\rm Im}\omega$ is small. 
Thus, if the right hand side is small, the equation cannot be satisfied.
At small values of $\omega$ and
$|{\bf k}|$ summation in the Eq.(\ref{dsp_bad}) can be substituted by
integration. Assuming that ${\rm Re}\omega \gg {\rm Im}\omega$ one
gets
\begin{equation}
\label{pl}
1 + \frac{ k^2 c_0^2}{\omega^2}=-i \frac{\sigma_{ef} f}{\omega
\epsilon_0}.
\end{equation}
This equation does not have real solutions for $\omega (k)$.  Thus, at
negative $\mu$ there are no propagating modes at any frequency under
the study.

This result obviously follows from the fact that $\mu \epsilon_0 <0$ in
the space between the wires. Therefore we get the plus sign in the left-hand side
of Eq.(\ref{pl}) which forbids any EMW propagation. Thus, instead of the LHM 
we get a material without any propagating modes.

Now we compare this result with the theoretical idea\cite{p2,sm} to obtain the LHM,
where negative $\epsilon$ is created by the system of wires and 
negative $\mu$ is created in some other way.
This idea is based upon the assumption that the 
negative $\epsilon$ at $\omega< \omega_0$ results from a ``longitudinal plasma mode".
It is taken for granted that its frequency 
is independent of magnetic properties of the system, which is 
usually the case for plasmons.
However, the mode discussed above is not a plasma mode (see also \cite{pcom}). 
One can show that
this mode has zero
average value of the magnetic induction 
$\overline{{\bf B}}$ over the unit cell.
In this sense this is indeed a longitudinal mode.
But the average
value of the magnetic energy, which is 
proportional to $\overline{{\bf B}^2}$, is not zero and it is large.   
The physics of this mode is
substantially related to the magnetic energy.  That is why
negative $\mu$ completely destroys  this mode. It destroys also 
the region of negative $\epsilon$. In fact 
 this could  be
predicted
from the observation that $\omega_0 \sim c/d$ becomes
imaginary at negative $\mu$.
Say, one can see from Eqs.(\ref{eps},\ref{om}) that at $\mu<0$ one gets
$\omega_p^2<0$ and $\epsilon>0$ at all frequencies assuming that $\Gamma$ is small.

Thus, we have shown that the simple explanation \cite{p2,sm} 
of the negative refraction
in  the compound system of the MPC and SRR's, 
based upon the  permittivity $\epsilon$ of the MPC and the negative permeability $\mu$ 
of the SRR's does not work because negative $\mu$ blocks propagation of EMW's  in the MPC. 
The propagation observed by the San Diego group might be a manifestation of the remarkable conclusion
 of Landau and Lifshitz (See Ref.\cite{lan} p.268) that $\mu (\omega)$ does not have physical
meaning starting with some low frequency. Then, 
the explanation of the negative refraction in this particular 
system would be outside the simple Veselago scenario (see Ref.\cite{notomi} as an example). 
In this case, to explain the negative refraction one should use a 
microscopic equation similar to Eq.(\ref{dsp}) 
but with the SRR's included.

Finally, an analytical theory of the low frequency EMW in the
two-dimensional MPC is proposed.
It is shown that
the propagation of the low frequency waves is possible
because electric and magnetic fields in the wave are strongly
inhomogeneous inside  the lattice cell
and that electric field is small near the wires.
If the dielectric part of the MPC has a
negative $\mu$, no waves can
propagate through the system. We argue that
 the explanation of the experiment
Ref.\cite{sm3} is much
deeper than it has been supposed before.

We are grateful to A. Zakhidov and V. Vardeny who have attracted our
attention to this interesting problem and to L. P. Pitaevskii and  B. L. Spivak
for important discussions. The work has been funded by the NSF grant DMR-0102964.

\end{document}